\documentstyle[aps,eqsecnum,preprint]{revtex}
\begin{document}

\title{Classical quasiparticle dynamics and chaos in trapped Bose
condensates\thanks{dedicated to Boris Chirikov on the occasion
of his 70$^{\rm th}$ birthday}}
\author{Martin Fliesser and Robert Graham}
\address{Fachbereich Physik, Universit\"at-Gesamthochschule Essen\\
45117 Essen\\ Germany}
\maketitle

\begin{abstract}
In the short wavelength limit the Bogoliubov quasiparticles of trapped
Bose-Einstein condensates  can be described as classical particles
and antiparticles with dynamics in a mixed phase-space.
For anisotropic parabolic traps we determine the location of the
resonances and study the influence of the sharpness of the condensate
surface on the appearance of chaos as the energy of the quasiparticles
is lowered from values much larger than to values comparable with the
chemical potential.
\end{abstract}

\section{Introduction}

The achievement of Bose-Einstein condensation of clouds of magnetically
trapped alkali atoms \cite{anderson,bradley,davis} by evaporation cooling
to temperatures
in the 100 nano-Kelvin regime has revived the interest in the physics of
weakly interacting Bose-condensed gases. As is well-known the bulk
properties of such gases are rather well described by ideal gases of
quasi-particles, first derived microscopically by Bogoliubov \cite{bog},
which are collisionless phonons at long wavelength and approach free particles
at short wavelength. In the case of trapped condensates of repelling
atoms, to which we confine our discussion here, the quasiparticle
description retains its usefulness. However, trapped condensates are
spatially inhomogeneous. They form around the minima of the trapping
potential and can form rather sharp surfaces. The thickness of the
surface layer, given by the healing length \cite{baym}, can become very
small compared to the radius of the condensate. Quasiparticles may therefore
leave the condensate and reenter it after being reflected back by the trapping
potential. Therefore their dynamics are more complex than in
a homogeneous system. In Bogoliubov's approach single quasiparticles are
described by wavefunctions, which are solutions of a set of linear
wave equations. This is in complete analogy to the description of single
particles in quantum mechanics by Schr\"odinger's equation. We know
that it is extremely fruitful to examine the classical limit of the
Schr\"odinger equation, which gives all of classical physics. In the
same spirit we can examine the classical limit of the wave equations for
the quasiparticles. In spatially homogeneous condensates this gives a
simple theory of free quasiparticles with conserved momentum which are
distinguished from free particles merely by their unusual relation between
energy and momentum, and hence also between velocity and momentum.

In the case of trapped inhomogeneous condensates recent work has shown
\cite{fliesser} that the classical limit of the dynamics of quasiparticles
becomes more interesting because they experience forces both from
the trap and from the condensate. In the case of isotropic traps, and
hence also isotropic condensates, angular momentum conservation ensures
the integrability of the quasiparticle dynamics. This can be used to
construct WKB solutions of Bogoliubov's wave equations for this case
\cite{csordas}. In the experimentally more relevant case of axially symmetric
traps integrability of the quasiparticle dynamics is lost and numerical
studies \cite{fliesser1,fliesser} indeed have shown a generally mixed
phase-space in this case.
Detailed analytical work was performed for quasiparticle energies
$E$ much smaller than the chemical potential $\mu$. Here two integrable
limits were identified: (i) the phonon limit, where the phonon-like
quasiparticle is confined to the interior of the condensate and is
specularly reflected back when it strikes the surface, and (ii)
a surface-particle limit, where the motion of the single-atom-like
quasiparticle consists of rapid small-amplitude oscillations between the
repelling main bulk of the condensate and the potential wall of the
trap and a slow secular motion along the surface of the condensate. The
numerical examination of the quasiparticle dynamics at larger energies
(and for the trap anisotropy  of the experiment \cite{anderson}) revealed
\cite{fliesser} a strong chaotic component at $E=\mu$ and a curious
`quasi-integrable' regime at $E\gg \mu$, where the dynamics in phase-space
clusters around the tori corresponding to single atom motion in the
trapping potential, however with small-scale chaos superimposed.

In the present paper we wish to study this large energy regime more
closely, both analytically and numerically. In the spirit of Chirikov's
pioneering work on the onset of chaos (reviewed in \cite{chirikov}) we ask
`where are the resonances'
and answer this question by developing the first steps of the classical
perturbation theory for the quasiparticle dynamics at large energy. We
also examine the influence of the thickness of the surface on the
appearance of chaos. In the Thomas-Fermi approximation \cite{baym} this
thickness
is neglected, leading to a spatially discontinuous effective force on the
quasiparticle. This violates assumptions of the KAM theorem and turns out
to be the reason that noticeable small-scale chaos survives even at very
large energies, where the system should be close to the integrable
limit of independent atoms in the trap. Simulations with different boundary
layers substantuate this hypothesis. For a condensate with boundary layer
we study the transition to chaos as the energy is lowered to values comparable
with the chemical potential.

\section{Equations of motion}

In the present section we give a brief derivation of the relevant
equations of motion. The starting point is the Gross-Pitaevskii equation
\cite{gross}
\begin{equation}
\label{eq:2.1}
 i\hbar\dot{\psi}(\bbox{x},t)=
  \left\{-\frac{\hbar^2}{2m}\nabla^2+U(\bbox{x})+V_0|\psi(\bbox{x},t)|^2
  \right\}\psi(\bbox{x},t)
\end{equation}
describing the macroscopic wavefunction $\psi$ of a Bose condensate at
temperature $T=0$ in a trap with potential $U$ and interaction
$V_0=4\pi\hbar^2 a/m$, where $a$ is the $s$-wave scattering length. We
shall consider only the case of repulsive 2-particle interactions $a>0$.
The trap potential is taken as axially symmetric and parabolic
\begin{equation}
\label{eq:2.2}
 U(\bbox{x})=\frac{m\omega^2_0}{2}(x^2+y^2+\lambda z^2)\,.
\end{equation}
The ratio between axial and radial trap frequency is $\sqrt{\lambda}$.
The particle number $N=\int d^3\!x\,|\psi(\bbox{x},t)|^2$ is conserved and
assumed fixed. The equilibrium state of the condensate with chemical
potential $\mu$ is a solution of (\ref{eq:2.1}) with
$\psi(\bbox{x},t)=e^{-i\mu t/\hbar}\psi_0(\bbox{x})$ which satisfies
\begin{equation}
\label{eq:2.3}
 \mu\psi_0(\bbox{x})=
  \left\{-\frac{\hbar^2\nabla^2}{2m}+U(\bbox{x})+V_0|\psi_0(\bbox{x},t)|^2
  \right\}\psi_0(\bbox{x})
\end{equation}
with vanishing axial angular momentum
\begin{equation}
\label{eq:2.4}
 L_z=\frac{\hbar}{i}\int d^3\!x\,\psi^*_0(\bbox{x})
  (\bbox{x}\times\bbox{\nabla})\psi_0(\bbox{x})\,.
\end{equation}
The small oscillations around the equilibrium state are waves of the form
\begin{equation}
\label{eq:2.5}
 \psi(\bbox{x},t)=e^{-i\mu t/\hbar}
  \left(\psi_0(\bbox{x})+\varphi(\bbox{x},t)\right)
\end{equation}
with $\varphi$ considered as small and satisfy a linearized version of
eq.~(\ref{eq:2.1})
\begin{equation}
\label{eq:2.6}
 i\hbar\varphi(\bbox{x},t)=
  \left\{-\frac{\hbar^2}{2m}\nabla^2+U(\bbox{x})-\mu+2V_0
   |\psi_0(\bbox{x})|^2\right\}\varphi(\bbox{x},t)
 + V_0\psi^2_0(\bbox{x})\varphi^*(\bbox{x},t)\,.
\end{equation}
Usually the small oscillations are considered in second quantization (where
(\ref{eq:2.6}) becomes an operator equation, but retains its form), which
brings out their particle-like properties, and they are then called
quasiparticles. Here we shall bring out the particle properties in
a different way, by considering the `ray-optics' limit of eq.~(\ref{eq:2.6})
\cite{csordas}.
Formally, this is done by means of the ansatz
\begin{equation}
\label{eq:2.7}
 \varphi(\bbox{x},t)=\left(a_0(\bbox{x},t)+O(\hbar)\right)
  e^{iS(\bbox{x},t)/\hbar}-\left(b_0(\bbox{x},t)+O(\hbar)\right)
   e^{-iS(\bbox{x},t)/\hbar}
\end{equation}
and considering the terms arising from eq.~(\ref{eq:2.6}) formally order
by order in $\hbar$. The symmetrical form of the ansatz (\ref{eq:2.7})
allows us to restrict the sign of $\partial S/\partial t$ by choosing
$\partial S/\partial t=-E<0$. $a_0$ and $b_0$ can be interpreted as
semiclassical amplitudes of the quasiparticles and their antiparticles,
repectively. Writing $\bbox{p}=\bbox{\nabla}S$ for their momentum one finds
to zeroth order coupled algebraic equations for the amplitudes $a_0$, $b_0$
\begin{equation}
\label{eq:2.8}
 \left(\begin{array}{cc}
        \epsilon_{HF}(\bbox{x},\bbox{p})+
	 \frac{\partial S(\bbox{x},\bbox{t})}{\partial t} &
	  -V_0\psi^2_0(\bbox{x})\\
	-V_0\psi^{*2}_0(\bbox{x}) & \epsilon_{HF}(\bbox{x},\bbox{p})-
	 \frac{\partial S(\bbox{x},\bbox{t})}{\partial t}\end{array}
\right)
\left(\begin{array}{c}
       a_0(\bbox{x},\bbox{t})\\ b_0(\bbox{x},\bbox{t})\end{array}
\right)=0
\end{equation}
where
\begin{equation}
\label{eq:2.9}
 \epsilon_{HF}=\frac{p^2}{2m}+U(\bbox{x})-\mu+2V_0|\psi_0(\bbox{x})|^2
\end{equation}
is called the classical Hartree-Fock energy.

Eq.~(\ref{eq:2.8}) implies as solvability condition a quadratic equation
for $\partial S/\partial t$ which reduces to the Hamilton-Jacobi equation
\begin{equation}
\label{eq:2.10}
 \frac{\partial S}{\partial t}+H(\bbox{x},\bbox{\nabla} S)=0
\end{equation}
with
\begin{equation}
\label{eq:2.11}
 H(\bbox{x},\bbox{p})=
  \sqrt{\epsilon_{HF}^2(\bbox{x},\bbox{p})-V_0^2|\psi_0(\bbox{x})|^4}
\end{equation}
where by the sign convention on $\partial S/\partial t$ the positive branch
of the square-root must be taken. Eqs.~(\ref{eq:2.10}), (\ref{eq:2.11})
now describe the classical dynamics of the quasiparticles complementary
to the waves with dynamics (\ref{eq:2.6}). The conservation of energy
$H$ and axial angular momentum $L_z$ of the quasiparticles can be taken
care off by separating
\begin{equation}
\label{eq:2.12}
 S(\bbox{x},t) =S_0(\rho,z)-Et-L_z\phi
\end{equation}
where $\rho$, $\phi$, $z$ are standard cylinder coordinates.
Eq.~(\ref{eq:2.10}) then reduces to
\begin{equation}
\label{eq:2.13}
 H\left(\rho,z,\frac{\partial S_0}{\partial\rho}\,,
  \frac{\partial S_0}{\partial z}\right)=E
\end{equation}
with $|\psi_0(\bbox{x})|^2=|\psi_0(\rho,z)|^2$ and $p^2=p^2_\rho+p^2_z
+L^2_z/\rho^2$ in eqs.~(\ref{eq:2.9}), (\ref{eq:2.11}). Eq.~(\ref{eq:2.8})
fixes only the ratio of the amplitudes $b_0$ and $a_0$, which,
for fixed $E$ and $L_z$, becomes
\begin{equation}
\label{eq:2.14}
 b_0=\left[
  \frac{(E^2+V_0^2|\psi_0(\bbox{x})|^4)^{1/2}-E}
       {(E^2+V_0^2|\psi_0(\bbox{x})|^4)^{1/2}+E}\right]^{1/2}a_0\,.
\end{equation}
In order to determine the absolute values (apart from an arbitrary
space-independent factor) one has to go to the next order in the expansion
\cite{fliesser} where one finds as a solvability condition the conservation
law
\begin{equation}
\label{eq:2.15}
 \frac{\partial}{\partial t}(|a_0|^2-|b_0|^2)+\frac{1}{2m}
  \bbox{\nabla}\cdot(|a_0|^2+|b_0|^2)\bbox{\nabla}S=0\,.
\end{equation}
In the following we shall be concerned only with the classical quasiparticle
dynamics described by eq.~(\ref{eq:2.13}) with the Hamiltonian
(\ref{eq:2.11}), (\ref{eq:2.9}).

\section{Perturbation theory for large energy}

The single-particle interaction energy with the condensate is of the order
of the chemical potential $\mu$ and is only a small perturbation for energies
$E\gg\mu$. Expanding to second order the Hamiltonian takes the form
\begin{equation}
\label{eq:3.1}
 H=\epsilon_0(\bbox{x},\bbox{p})+2V_0|\psi_0(\bbox{x})|^2-
  \frac{V_0^2}{2}\,\frac{|\psi_0(\bbox{x})|^4}{\epsilon_0(\bbox{x},\bbox{p})}
\end{equation}
with
\begin{equation}
\label{eq:3.2}
 \epsilon_0=\frac{\bbox{p}^2}{2m}+\frac{m}{2}
  \omega_0^2(x^2+y^2+\lambda z^2)-\mu\,.
\end{equation}
In the following we shall restrict our discussion mainly
to the case of vanishing
axial angular momentum $L_z=0$,
but at the end we also present some results for $L_z\ne0$
in order to assess to what extent the case $L_z = 0$
already captures the typical behaviour of the system.
Experimentally modes with  $L_z=0$ or $L_z\ne0$
can be excited depending on the symmetry of the excitation mechanism.

For $L_z=0$ the dynamics is restricted to a plane containing
the $z$-axis, which can be taken as the $(x,z)$-plane $y\equiv 0$, $p_y=0$.
The action angle variables of the unperturbed harmonic motion in the trap are
\begin{eqnarray}
\label{eq:3.3}
 x&=&\sqrt{\frac{2{\rm I}_x}{m\omega_0}}\sin\theta_x\phantom{\sqrt{\lambda}}
   \quad,\quad
  p_x=\sqrt{2m\omega_0{\rm I}_x}\cos\theta_x\nonumber\\
  &&\\
 z&=&\sqrt{\frac{2{\rm I}_z}{m\sqrt{\lambda}\omega_0}}\sin\theta_z
   \quad,\quad
  p_z=\sqrt{2m\sqrt{\lambda}\omega_0{\rm I}_z}\cos\theta_z\nonumber
\end{eqnarray}
with
\begin{equation}
\label{eq:3.4}
 \epsilon_0=\omega_0{\rm I}_x+\sqrt{\lambda}\omega_0{\rm I}_z-\mu\,.
\end{equation}
To express the perturbed Hamiltonian in action-angle variables we need
the Fourier coefficients of the condensate density and its square
\begin{eqnarray}
\label{eq:3.5}
 \bar{\rho}_{\ell n}({\rm I}_x,{\rm I}_z) &=&
  \frac{1}{(2\pi)^2}\int_0^{2\pi}d\theta_x\int_0^{2\pi}d\theta_z
   e^{-i(\ell\theta_x+n\theta_z)}|\psi_0(x,0,z)|^2\nonumber\\
 &&\\
 \bar{\rho^2}_{\ell n}({\rm I}_x,{\rm I}_z) &=&
   \frac{1}{(2\pi)^2}\int_0^{2\pi}d\theta_x\int_0^{2\pi}d\theta_z
    e^{-i(\ell\theta_x+n\theta_z)}|\psi_0(x,0,z)|^4\nonumber\,.
\end{eqnarray}
The canonical transformation  $(\theta,{\rm I})\to(\phi,J)$ with
\begin{eqnarray}
\label{eq:3.6}
 {\rm I}_x &=& J_x-\frac{2V_0}{\omega_0}\sum_{\ell,n}
  \frac{\ell}{\ell+\sqrt{\lambda}n}\,\bar{\rho}_{\ell,n}(J_x,J_z)
   e^{i(\ell\theta_x+n\theta_z)}\nonumber\\
 &&\\
 \phi_x &=& \theta_x-\frac{2V_0}{\omega_0}\sum_{\ell,n}
  \frac{1}{i(\ell+\sqrt{\lambda}n)}\,
   \frac{\partial\bar{\rho}_{\ell,n}(J_x,J_z)}{\partial J_x}
    e^{i(\ell\theta_x+n\theta_z)}\nonumber
\end{eqnarray}
and analogous for ${\rm I}_z$, $\phi_z$ removes the angle-dependence in
the first order perturbation term and we are left with the second-order
Hamiltonian
\begin{equation}
\label{eq:3.7}
 H=\omega_0(J_x+\sqrt{\lambda}J_z)-\mu+2V_0\bar{\rho}_{00}(J_x,J_z)+
  \frac{V_0^2}{\omega_0}\sum_{\ell,n}K_{\ell,n}(J_x,J_z)
   e^{i(\ell\phi_x+n\phi_z)}
\end{equation}
where
\begin{equation}
\label{eq:3.8}
 K_{\ell,n}=-
  \frac{\bar{\rho^2}_{\ell,n}}{2(J_x+\sqrt{\lambda}J_z)-2\mu/\omega_0}-
  4\sum_{p,r}
   \left(p\frac{\partial\bar{\rho}_{\ell-p,n-r}}{\partial J_x}+
    r\frac{\partial\bar{\rho}_{\ell-p,n-r}}{\partial J_z}\right)
     \frac{\bar{\rho}_{p,r}}{p+r\sqrt{\lambda}}\,.
\end{equation}
Eq.~(\ref{eq:3.7}) differs from (\ref{eq:3.1}) only by terms of higher
than second order. For irrational $\sqrt{\lambda}$ there are no
resonances to zeroth order in the interaction, but in first order isolated
resonances $\Omega_x(J_x,J_z)P+\Omega_z(J_x,J_z)R=0$ appear with integers
$P$, $R$ and the perturbed frequencies
\begin{eqnarray}
\label{eq:3.9}
 \Omega_x &=& \omega_0+2V_0\frac{\partial\bar{\rho}_{00}}{\partial J_x}
 \nonumber\\
 &&\\
 \Omega_z &=& \sqrt{\lambda}\omega_0+2V_0
  \frac{\partial\bar{\rho}_{00}}{\partial J_z}\nonumber\,.
\end{eqnarray}
Their distances in frequency space have to be compared with their widths,
given by \cite{chirikov}
\begin{equation}
\label{eq:3.9a}
 \Delta\omega= 4\sqrt{\frac{V_0^3}{\omega_0}\,
  \frac{|K_{PR}(J_x,J_z)|}{M_{PR}(J_x,J_z)}}
\end{equation}
where
\begin{equation}
\label{eq:3.9b}
 \frac{1}{M_{PR}(J_x,J_z)}=\frac{\sum_{i,k=x}^zP_i
  \frac{\partial^2\bar{\rho}_{00}}{\partial J_i\partial
J_k}P_k}{P_x^2+P_z^2},\,\quad
   P_x=P,\quad P_z=R\,.
\end{equation}
If resonances
overlap somewhere in phase space Chirikov's criterion \cite{chirikov} tells
us that we should
expect chaos there. Unfortunately, eqs.~(\ref{eq:3.7}), (\ref{eq:3.8}) are
too difficult to evaluate analytically for a realistic equilibrium solution
$\psi_0(\bbox{x})$ of the time-independent Gross-Pitaevskii equation.
However, these equations are still useful to understand some qualitative
features of numerical simulations of the complete Hamiltonian dynamics
at large energies.

Let us discuss in particular the case of large condensates to which the
Thomas-Fermi approximation applies, in which the spatial derivative terms
in eq.~(\ref{eq:2.3}) are neglected. The condensate density is then given
by
\begin{equation}
\label{eq:3.10}
 |\psi_0(\bbox{x})|^2=
  \frac{\mu-(m\omega_0^2/2)(x^2+y^2+\lambda z^2)}{V_0}
   \theta (\mu-\frac{m\omega_0^2}{2}(x^2+y^2+\lambda z^2))\,.
\end{equation}
where $\theta (x)$ is the Heaviside step-function. As $|\psi_0(\bbox{x})|^2$
is an even function of $x$, $y$, $z$ the perturbation amplitudes are
nonvanishing only for even $\ell,n$. Therefore  $K_{PR}$ in (\ref{eq:3.9a})
must be replaced by $K_{2P2R}$. The Thomas-Fermi
approximation of the condensate density (\ref{eq:3.10}) has a discontinuous
first-order derivative at the surface. This means that the
Fourier coefficients $\bar{\rho}_{\ell,n}$ and $\bar{\rho^2}_{\ell,n}$
at large $|\ell|$, $|n|$ fall off like $|\ell|^{-2}$, $|n|^{-2}$ and
$|\ell|^{-3}$, $|n|^{-3}$, respectively giving the $(P,R)$- resonances
for large $|P|$, $|R|$ widths, which according to eq.~(\ref{eq:3.8}), fall
off only like $|P|^{-1}$, $|R|^{-1}$. This estimate results from the second
term in eq.~(\ref{eq:3.8}), which is predicted to fall off only like
$|\ell|^{-2}$, $|n|^{-2}$. On the other hand the number of
large-order resonances $\Omega_x/\Omega_z=-R/P$ scales like $|P|$,
$|R|$. Therefore, barring non-generic cases where the $|K_{PR}|$ are small
for some exceptional reason, one expects large-order resonances to
always overlap in the Thomas-Fermi approximation, i.e. the tori of the
free harmonic oscillations in the trap will typically all be broken.
By the same arguments a condensate density with $M$ smooth derivatives
and a discontinuous $M+1$ order derivative will give rise to resonance
widths scaling like $|P|^{-\frac{M}{2}-1}$, $|R|^{-\frac{M}{2}-1}$ and
tori with large $|P|$, $|R|$ can exist if $M>M_c=2$. The critical order
of smoothness for general Hamiltonian systems with $f$ degrees of freedom
was determined  by Chirikov \cite{chirikov} as $M_c=2f-2$.

In fig.~1 a Poincar\'e surface of section of the dynamics at $E=20\mu$,
$L_z=0$, taken at $x=0$ and plotted in the $(\theta_z,{\rm I}_z)$-plane
is shown for the experimentally realized \cite{anderson} value $\lambda=8$.
The action variable ${\rm I}_z$ is plotted in units of $2\mu/\omega_0$.
With the exception of tori at small ${\rm I}_z$ and at large
${\rm I}_z\simeq (E+\mu)/\sqrt{\lambda}\omega_0$ all tori in fig.~1 are
broken. The survival of the exceptional unbroken tori can be understood
from the fact that for $I_z \to 0$ or $I_x \to 0$ (with
$I_z\simeq (E+\mu)/\sqrt{\lambda}\omega_0$) all the coefficients
$K_{\ell,n}$ approach $0$ with the exception of
$K_{\ell,0}$ or $K_{0,n}$, repectively, which cannot
influence appreciably tori with frequency ratios
$\Omega_x/\Omega_z\simeq8^{-1/2}$.

In fig.~1 a number of resonances with frequency ratios in the neighborhood
of $\Omega_x/\Omega_z=8^{-1/2}$ can be discerned. Their frequency ratios
are given on the right hand side of the graph.
To understand the position where these resonances occur
we consider $\bar{\rho}_{00}$ in the Thomas-Fermi approximation
\begin{eqnarray}
\label{eq:3.11}
 \bar{\rho}_{00} = \frac{4}{\pi^2V_0}\int_0^{\frac{\pi}{2}}d\phi_x
  \int_0^{\frac{\pi}{2}}d\phi_z
   &&\left(\mu-\omega_0 J_x\sin^2\phi_x-\sqrt{\lambda}\omega_0 J_z\sin^2
     \phi_z\right)\nonumber\\
   && \quad\theta\left(\mu-\omega_0 J_x\sin^2\phi_x-\sqrt{\lambda}
       \omega_0 J_z\sin^2\phi_z\right)
\end{eqnarray}
and evaluate the first-order frequency shifts
\begin{equation}
\label{eq:3.12}
 \Delta\omega_{x,z}=2V_0\frac{\partial\rho_{00}}{\partial J_{x,z}}
\end{equation}
from the integrals
\begin{eqnarray}
\label{eq:3.13}
 \Delta\omega_x &=& -\frac{8\omega_0}{\pi^2}\int_0^{\pi/2} d\phi_x
  \int_0^{\pi/2} d\phi_z\sin^2\phi_x\theta
   (\mu-\omega_0 J_x\sin^2\phi_x-\sqrt{\lambda}\omega_0 J_z\sin^2\phi_z)
 \nonumber\\
  &&\\
   \Delta\omega_z &=& -\frac{8\omega_0\sqrt{\lambda}}{\pi^2}
    \int_0^{\pi/2} d\phi_x\int_0^{\pi/2} d\phi_z\sin^2\phi_z\theta
     (\mu-\omega_0 J_x\sin^2\phi_x-\sqrt{\lambda}\omega_0 J_z\sin^2\phi_z)
     \nonumber\,.
\end{eqnarray}
It is manifest that $\Delta\omega_{x,z}$ are negative. Near the bottom of
fig.~1 the action $J_z$ is small, while
$J_x\simeq (E+\mu)/\omega_0-\sqrt{\lambda} J_z$ is large. The Heaviside
function in eqs.~(\ref{eq:3.13}) therefore restricts $\sin^2\phi_x$ to small
values of the order of $\mu/\omega_0 J_x=O(\mu/E)$ while
$\sqrt{\lambda}\sin^2\phi_z$ is not so restricted. Therefore
$|\Delta\omega_x|\ll|\Delta\omega_z|$ near the bottom of fig.~1 and
$ \frac{\Omega_x}{\Omega_z}-\frac{1}{\sqrt{\lambda}}\simeq
\frac{\Delta\omega_x}{\sqrt{\lambda}\omega_0}-
\frac{\Delta\omega_z}{\omega_0\lambda}>0$
holds there. In the upper parts of fig.~1 $J_x$ is small while
$J_z\simeq (E+\mu)/\sqrt{\lambda}\omega_0-J_x/\sqrt{\lambda}$. The situation
is therefore reversed and we obtain
$\frac{\Omega_x}{\Omega_z}-\frac{1}{\sqrt{\lambda}}< 0$ by the same argument.

The periodic orbit $z=0$, $p_z=0$ which has $J_z=0$ forms the lower
border of the range of $J_z$. For this case the
frequency shifts $\Delta\omega_x$, $\Delta\omega_z$ are easily evaluated
from eq.~(\ref{eq:3.13}) with the result
\begin{eqnarray}
\label{eq:3.15}
 \Delta\omega_x &=& -O\left(\omega_0 (\mu/\omega_0 J_x)^{3/2}\right)
  \nonumber\\
  &&\\
  \Delta\omega_z &=& -\frac{2}{\pi}\sqrt{\frac{\lambda\omega_0\mu}{J_x}}
  \nonumber
\end{eqnarray}
indicating that the fixed point is stable for $E\gg\mu$. The arguments of
the previous paragraph indicate, furthermore, that $|\Delta\omega_x|$ is
minimal for this case, while $|\Delta\omega_z|$ is maximal, leading to
a maximal value of the ratio
\begin{equation}
\label{eq:3.16}
 \left(\frac{\Omega_x}{\Omega_z}\right)_{\max}\simeq\frac{1}{\sqrt{\lambda}}
  +\frac{2}{\pi}\sqrt{\frac{\mu}{\lambda\omega_0 J_x}}\simeq
    \frac{1}{\sqrt{\lambda}}
  \left(1+\frac{2}{\pi}\sqrt{\frac{\mu}{E+\mu}}\right)
\end{equation}
where we used $E+\mu\simeq\omega_0 J_x+\sqrt{\lambda}\omega_0J_z$ with
$J_z=0$ in
the last estimate. For $E/\mu=20$ this gives
$(\Omega_x/\Omega_z)\simeq0.402$, which is just barely larger than the
ratio 0.4 of the 2:5 resonance visible near the lower border of fig.~1.
Similar arguments apply to the upper border of the range of $J_z$ which is
formed by the periodic orbit $x=0$, $p_x=0$ with vanishing $J_x$ for which
\begin{equation}
\label{eq:3.17}
 \Delta\omega_x =-\frac{2}{\pi}\sqrt{\frac{\omega_0\mu}{\sqrt{\lambda}J_z}}
 \quad,\quad
  \Delta\omega_z=-\sqrt{\lambda\omega_0}O\left((\mu/\sqrt{\lambda}
  \omega_0J_z)^{3/2}\right)\,.
\end{equation}
Thus
\begin{equation}
\label{eq:3.18}
 \left(\frac{\Omega_x}{\Omega_z}\right)_{\min}\simeq\frac{1}{\sqrt{\lambda}}
  \left(1-\frac{2}{\pi}\sqrt{\frac{\mu}{E+\mu}}\right)\,.
\end{equation}
For $E/\mu=20$ this yields $(\Omega_x/\Omega_z)_{\min}\simeq 0.304$ which is
smaller than the ratio 0.333 of the 1:3 resonance visible near the upper
border of fig.~1.

In general we can conclude that for fixed energy $E$ we should expect to see
the strongest resonances in the interval
\begin{equation}
\label{eq:3.19}
 \frac{1}{\sqrt{\lambda}}
  \left(1-\frac{2}{\pi}\sqrt{\frac{\mu}{E+\mu}}\right)<\Omega_x/\Omega_z<
   \frac{1}{\sqrt{\lambda}}
    \left(1+\frac{2}{\pi}\sqrt{\frac{\mu}{E+\mu}}\right)\,,
\end{equation}
which are, for $\lambda=8$, $E/\mu=20$, the resonances 1:3, 2:5, 3:8, 4:11,
5:14, 6:17 and are indeed all seen in fig.~1 with the exception of the 5:14
resonance. In its place a chaotic region is seen. The resonance values of
the actions following from the perturbation theory are indicated on the
right-hand side of fig.1. The 4:11 resonance predicted close to the missing
5:14 resonance is nearly completely destroyed and barely separated from the
chaotic region replacing the 5:14 resonance.

Let us now consider the effect of the sharpness of the boundary layer of the
condensate. Its thickness is given by the healing length \cite{baym}
\begin{equation}
\label{eq:3.20}
 \ell_H=\frac{\hbar}{\sqrt{2\mu m}}\,.
\end{equation}
It enters as an independent parameter whose ratio to the radial Thomas-Fermi
radius $\rho_{TF}=\sqrt{2\mu/m\omega_0^2}$ is determined by the value of the
chemical potential, i.e. by the number of particles in the trap,
$\ell_H/\rho_{TF}=\hbar\omega_0/2\mu$. As example we consider
$\ell_H/\rho_{TF}=0.1$. The condensate with
boundary layer is modelled by joining
\begin{eqnarray}
\label{eq:3.21}
 |\psi_0(\bbox{x})|^2 &=& \frac{\mu-\frac{1}{2}m\omega_0^2r^2}{V_0}
  \theta(\rho_{TF}-\ell_H-r)\nonumber\\
 && +\exp(-a_0-a_1(r-\rho_{TF}+\ell_H)-a_2(r-\rho_{TF}+\ell_H)^2\\
&& \qquad\quad -a^3(r-\rho_{TF}+\ell_H)^3)\theta(r-\rho_{TF}+\ell_H)
\nonumber
\end{eqnarray}
continuously with continuous three derivatives at $r=\rho_{TF}-\ell_H$,
thereby fixing $a_0,\dots,a_3$. Here the abreviation
$r=\sqrt{\rho^2+\lambda z^2}$ is used. The degree of smoothness thereby
introduced should be sufficient to see smooth KAM tori. The model
condensate (\ref{eq:3.21}) is of course not
a solution  of the Gross-Pitaevskii equation, but does not differ
qualitatively from solutions taking into account the boundary layer
\cite{dalfovo}. For our present purposes it is therefore perfectly acceptable
while avoiding unnecessary complications. In fig.~2 we compare
Poincar\'e surface of sections in the same format as in fig.1 for the
Thomas-Fermi approximation and the smooth model condensate with
$l_H/\rho_{TF}=0.1$ at energy $E/\mu=100$ and $\lambda=8$. It can be seen that
with the smooth boundary layer included most of the broken tori of the
Thomas-Fermi condensate seen in the upper part of fig.2 become smooth, as
seen in the lower part of fig.2, however with ripples on the tori still
indicating the presence of the narrow boundary layer. If the boundary layer is
narrowed further (not shown) these ripples become stronger and develop sharp
cusps. The chaotic band at small actions $I_z$ survives even for very
large quasiparticle energies $E/\mu$. Resonances within this band have
a comparatively large width for two reasons. For one the factors $1/M_{PR}$
are large, because $I_z$ is not far above the value
$I_z=\mu/\omega_0\sqrt{\lambda}$ where $M_{PR}^{-1}$ peaks (with a logarithmic singularity in Thomas-Fermi approximation). For larger
values of $I_z$, of the order of $E/\omega_0\sqrt{\lambda}$, like
for the 6:17 resonance or values of $P:R$ close to $\sqrt{\lambda}$,
$1/M_{PR}$ is much smaller by a factor of the order
of $(\mu/E)^2$.
Second, the interaction coefficient $|K_{PR}|$ has a resonantly enhanced
contribution $-4(\partial\bar{\rho}_{00}/\partial
J_z)\bar{\rho}_{2P2R}(P/R+\sqrt{\lambda})^{-1}$. The last factor in this
expression favors resonances $P:R$ close to the ratio $\sqrt{\lambda}$. The
existence of the narrow boundary layer of the condensate furthermore
leads to appreciable interaction coefficients $|K_{PR}|$ even for
comparatively high-order resonances whose overlap may be responsible for the
formation of the chaotic band. The requirement that {\it both} factors $M_{PR}^{-1}$ and $|K_{2P2R}|$ must be large may explain the appearance of a
band of actions $I_z$ with appreciable resonance overlap somewhat above the
value $I_z=\mu/\omega_0\sqrt{\lambda}$.

Dynamically the chaotic band is related to orbits in a layer of
actions $I_z\gtrsim \mu/\sqrt{\lambda}/\omega_0$ which are just sufficient for
the quasiparticle to hit or miss the condensate at random as it oscillates
back and forth in x-direction. This mechanism is able to introduce sensitive
dependence on initial conditions and exists, if only in a narrow band,
even at very large energies. For just slightly smaller actions
$I_z$ the quasiparticle has to pass the condensate twice in each period of
$\theta_x$ and the tori are smooth even in Thomas-Fermi approximation, as
can be seen in fig.~1 for $E/\mu=20$.
A similar mechanism for a chaotic band should actually exist also for the
oscillations in the z-direction, the short axis of the condensate-ellipsoid.
However the instabilty in this case for action
$I_x\gtrsim \mu/\sqrt{\lambda}/\omega_0$ seems to be much less pronounced
(which is plausible, because the perturbation by the condensate should be
weaker along the short axis) and is not visible in the numerical data.

Finally we consider numerically the transition to chaos in the condensate
with  boundary layer $l_H/\rho_{TF}=0.1$ as the quasiparticle energy is lowered.
In fig.3 the same Poincar\'e surface of section is plotted as in figs.1,2
(and again for $\lambda=8$) but in the $(z,p_z)$-plane rather than the
$(\theta_z,I_z)$-plane. The two plots in the upper row are for $E/\mu=100$
and $E/\mu=20$, those in the lower row are for $E=10$ and $E=2$ respectively.
The plots for $E=100$ and $E=20$ can be compared with fig.2 and fig.1
respectively, but fig.1 is, of course for vanishing $l_H$ only.
The slightly rounded Thomas-Fermi surface at $z/\rho_{TF}=\lambda^{-1/2}$
is visible in these graphs. Based on
these and similar plots the transition to chaos can now be roughly described
as follows: The chaotic band in a layer of small actions
$I_z\gtrsim \mu/\sqrt{\lambda}\omega_0$ existing even at very large energies
(see the plot for $E/\mu=100$) becomes wider as the energy is lowered
(see $E/\mu=20$). Then, at a critical energy $E/\mu=13.9$ which we discuss
further below, the periodic orbit along the long axis of the
condensate-ellipsoid becomes unstable and a second chaotic region in
phase-space surrounding the unstable orbit $z=0=p_z$ is created. The inner
and outer chaotic regions then rapidly grow together as the energy is
lowered further (see $E/\mu=10$) and finally fill up most of the accessible
regions of phase-space (see the last plot with $E/\mu=2$).

Apart from a continuous widening of the chaotic regions the main event in this
transition to chaos is the appearance of the instabilty of the periodic orbit
along the long axis of the condensate-ellipsoid, which for
$\lambda>0$ is the $x$-axis. In
Thomas-Fermi approximation this orbit has the time-dependence for
$|x|<\rho_{TF}$
\begin{eqnarray}
\label{eq:3.22}
 x &=&  \rho_{TF}\sqrt{E/\mu-1}\sinh (\omega_0t-\varphi^\circ)
  \nonumber\\
 &&\\
 p_x &=& m\omega_0\rho_{TF}\sqrt{E/\mu-1}\cosh (\omega_0t-\varphi^\circ)
 \nonumber
\end{eqnarray}
and for $|x|>\rho_{TF}$
\begin{eqnarray}
\label{eq:3.23}
 x &=&  \rho_{TF}\sqrt{E/\mu+1}\sin (\omega_0t-\varphi^{'\circ})
  \nonumber\\
 &&\\
 p_x &=& m\omega_0\rho_{TF}\sqrt{E/\mu+1}\cosh (\omega_0t-
 \varphi^{'\circ})
 \nonumber
\end{eqnarray}
with $\varphi^\circ$ and $\varphi^{'\circ}$ suitably adjusted by continuity at
$|x|=\rho_{TF}$.
To determine the energy where this orbit looses its stability we can perform
a linear stability analysis for small perturbations $z$, $p_z$ away from this
orbit, which satisfy
\begin{equation}
\label{eq:3.24}
 m\dot{z}= p_z,\quad
 \dot{p}_z = \lambda\left(1-2\theta(|x(t)|-\rho_{TF})\right)m\omega_0^2z\,.
\end{equation}
The growth of a perturbation of the periodic orbit is
determined by the monodromy matrix {\bf M} for a half-period $T/2$
\begin{equation}
\label{eq:3.25}
 \left(\begin{array}{l}
      z(T/2)/\rho_{TF}\\
      p_z(T/2)/m\omega_0\rho_{TF}\end{array}\right)=\bbox{M}
  \left(\begin{array}{l}
       z^{(0)}/\rho_{TF}\\
       p_z^{(0)}/m\omega_0\rho_{TF}\end{array}\right)\,.
\end{equation}
The Hamiltonian form of the dynamics ensures that ${\rm Det}~\bbox{M}=1$.
The stability condition for the perturbations in $z$-direction then becomes
\begin{equation}
\label{eq:3.26}
 |Tr\bbox{M}|\le2\,.
\end{equation}
With a little algebra it is straight-forward to evaluate {\bf M} and its
trace thereby reducing (\ref{eq:3.26}) to
\begin{equation}
\label{eq:3.27}
 |\cos\left(\sqrt{\lambda}\omega_0t_2\right)
 \cosh\left(\sqrt{\lambda}\omega_0t_1\right)|\le1
\end{equation}
Here $t_1$ and $t_2$ are the total lengths of the time-intervals during
each half-period where $|x|<\rho_{TF}$ and $|x|>\rho_{TF}$, respectively.
They are given by
\begin{equation}
\label{eq:3.28}
 t_1 = \frac{2}{\omega_0}\,\mbox{\rm Artanh}\,\sqrt{\frac{\mu}{E}},\qquad
  t_2 = \frac{2}{\omega_0}\,\mbox{\rm Arctan}\,\sqrt{\frac{E}{\mu}}
   \,.
\end{equation}
For $\lambda=8$ stability is lost, according to the criterion (\ref{eq:3.27})
for $E/\mu=14.4$ which is in reasonable agreement with the already
quoted value $E/\mu=13.9$ determined from the numerical simulation of the
complete quasiparticle dynamics.

\section{Conclusion}
In the present paper we have studied the classical limit of the dynamics
of the quasiparticles in a spatially inhomogeneous harmonically trapped
weakly interacting Bose gas.
As is well known many properties of a Bose gas are determined by its
quasiparticles. For sufficiently large energies $E/\hbar\omega_0\gg 1$ a
classical description of quasiparticles should be possible. We have
accordingly concentrated our attention on the large energy regime. For
quasiparticle energies large compared to the chemical potential a
perturbative expansion
becomes possible. The dynamics then looks like that of an anisotropic
harmonically bound particle which is perturbed by the presence of a weakly
repelling condensate localized around the center of the trap with a rather
sharp surface. The appearance of an infinitely sharp surface in the
Thomas-Fermi approximation leads to the break-up of all tori with the
exception of those in the immediate phase-space neighborhood of the
periodic orbits along the main axes of the ellipsoidal condensate. Therefore
in this case chaos exists in phase-space for arbitrarily large energies
$E/\mu$. Still resonances can be identified even in this case and their
actions are well described by perturbation theory. We have compared this
non-smooth case to that of a condensate with a smooth (up to third
derivatives) but narrow boundary layer. In real condensates the thickness of
the boundary layer is determined by the two-particle interaction and the
number of particles \cite{baym}. In that case smooth KAM tori exist at
sufficiently large energies,
but they are rippled by the influence of the boundary layer. Furthermore it
turns out that an appreciable  region of chaos in phase-space persists
even to large energies. It is related to orbits along the long axis of the
condensate-ellipsoid which are sufficiently perturbed in the direction of the
short axis to sometimes miss and sometimes hit the condensate in a random
way with sensitive dependence on small perturbations. Finally we have studied
how large scale chaos appears as the energy is gradually lowered
to values of the order of the chemical potential. The
instability of the periodic orbit along the long axis of the ellipsoid was
found to play a major role in this transition. It is connected with the
appearance of a second inner chaotic region at lower energies which joins
up with the chaotic band existing also at large energies.

Our discussion so far has been restricted to the special case $L_z = 0$,
where the classical quasiparticle dynamics is confined
to a plane in configuration space
containing the z-axis.
One may well ask to what extent this motion already captures the typical behavior of the system. To examine this question at least numerically
we have generated Poincar\'e surface of sections for $L_z\ne 0$
both for fixed $E$ and varying $L_z$
or for fixed $L_z$ and varying $E$.
The surfaces of section are always taken at the value $\rho = \rho_0$
where the effective potential in radial direction,
which includes the centrifugal barrier, has its minimum.
In fig.4 we present a series of Poincar\'e surfaces of sections
for $L_z$ fixed at a rather high value (in units of $\hbar$)
$L_z= \mu/\omega_0$
and the same values of $E$
and also the same thickness of the boundary layer
as in fig.3.

It is apparent that taking $L_z\ne0$ the mirror symmetry
with respect to the z-axis and the $p_z$-axis
which is present at $L_z=0$
is lost and replaced by a point symmetry
with respect to the origin $z=p_z=0$,
which is, of course, to be expected.
However, apart from this obvious difference
the qualitative behavior displayed in fig.4
is the same as in fig.3.
As was shown in \cite{fliesser} the energy and the angular momentum
must satisfy the inequalities
\begin{eqnarray}
 E+\mu > \omega_0 L_z \quad\mbox{if}\quad E>\mu \,\\
E > (\omega_0 L_z)^2/4\mu \quad\mbox{if}\quad E<\mu \,.
 \nonumber
\end{eqnarray}
For $E>\mu$ and $|L_z|> 2\mu/\omega_0$
there exists a domain
$(\omega_0 L_z)^2/4\mu > E > \omega_0 L_z - \mu $
where the motion of the quasiparticles occurs outside the condensate
in the Thomas-Fermi limit,
and is therefore integrable.
Thus an additional transition from nonintegrable to integrable dynamics occurs
for very high angular momentum $|L_z| > 2\mu/\omega_0$
as the energy is lowered from above to below $E=(\omega_0 L_z)^2/4\mu$.

Let us finally turn to the quantum mechanical implications
of the classical dynamics we have studied.
Such implications exist both for the wavefunctions and the energies of single quasiparticle states
which are of course closely connected.
As the classical analysis is done in phase space
it is most convenient for a discussion
to use the Husimi distribution
$Q(\rho,p_{\rho},z,p_z) = |<\alpha|\psi>|^2$
where $|\alpha>$ is a coherent state of the harmonic oscillators
defined by the free trap.
For each quantum state $|\psi>$ the function $Q$ is a positive
quasiprobability on the phase space,
which due to the overcompleteness of the $|\alpha>$,
contains the full information on $|\psi>$.

Let us now turn to the Poincar\'e surface of sections
displayed in fig.3 and discuss the corresponding
quantum states via their $Q$-functions:
In fig.3a the $Q$-functions at $\rho = 0$ will be spread out
along the tori with (in the limit $\hbar \omega_0/\mu \to 0$)
narrow peaks on the tori.
The number of such states is semiclassically given
by the phase space volume of the energy shell
of thickness $\Delta E$ in units
$(2\pi\hbar)^2$.
The chaotic layer visible in fig.3a may correspond to several quantum states
(depending on the contribution of this layer to the phase space volume
of the energy shell in units $(2\pi\hbar)^2$)
which are all spread out along the layer
but with wave functions oscillating wildly
in the corresponding configuration space
so as to satisfy the orthogonality condition.
If indeed several such states exist in an energy shell
of thickness $\Delta E$ their energies will tend to repell each other.
The energies corresponding to $Q$-functions centered on tori will show
no such repulsion.

Essentially the same discussion applies to fig.3b, 3c where the chaotic layer
occupies a larger fraction of the phase-space volume
and finally merges into a single domain.
In the case of fig.3d,
where essentially a single chaotic domain survives two different types of
quantum states are possible:
The first possibility is that the quantum states
are all spread out over the chaotic domain
with strong oscillations of their
wave functions in the corresponding domain of configuration space
to satisfy orthogonality
and accompanying strong level repulsion.
The second possibility is that the quantum states show dynamical
localization
with respect to the action variable $I_z$.
In this case the quantum states still localize around different values of $I_z$
due to a coherent interference effect akin to Anderson localization.
In this case the wave functions can be orthogonal without energy level repulsion.
Which of these two possibilities is realized depends on the strength
of the chaotic change of the action variable $I_z$.
If this change is sufficiently large,
then the first possibility
(extended states with level repulsion) will be realized.
If however $I_z$ changes sufficiently weakly in a diffusive way
such that its variance satisfies $<\Delta I_z^2> = Dt$
with a diffusion constant D,
then the second possibility may be realized.
Then a fundamental estimate \cite{localz}
can be made of the localization length $\xi_z$
of $I_z$ according to $\xi_z^2 \sim D\hbar\rho(E)$
where $\rho(E)$ is the density of states at energy $E$.
Dynamical localization then is predicted to occur as soon as
$\xi_z \ll W_z$ where $W_z$ gives the width of the chaotic domain
in $I_z$. Thus $D \ll W_z^2/\hbar \rho(E)$ is required.
At present no reliable estimate of $D$ exists,
either analytically or numerically,
so the question which of the two behaviors will occur in a given trap at energies $E\sim \mu$ must be left open here.
Numerical computations of quasiparticle energies have been performed
for certain trap parameters and avoided crossings have been seen
in the results for the regime $E\sim\mu$ (see e.g.\ \cite{you}),
but a systematic study of the level statistics has not yet been
performed for this case.

However,
the level repulsion typical for classically chaotic systems
has already been built into a quantum theory of damping of the
low-energy collective modes due to scattering of thermally excited
quasiparticles at energies of the order of the chemical potential \cite{damp}.
This should only be a beginning. If the precedence of mesoscopic systems is
any hint, we may expect that more properties of trapped Bose condensates will
turn out to be infiltrated by chaos in the future.

\section*{Acknowledgements}

The present work is an outgrowth of earlier work
\cite{fliesser,csordas,fliesser1} done in collaboration with
Andr\'as Csord\'as and P\'eter Sz\'epfalusy to whom we are gratefull for
sharing their insights.
This work has been supported by the Deutsche
Forschungsgemeinschaft through the Sonderforschungsbereich 237
``Unordnung und gro{\ss}e Fluktuationen''.

\section*{Figure captions}

\noindent
{\bf Fig.~1.} Poincar\'e surface of section of the quasiparticle dynamics. The
cut is taken at constant energy at $x=0$, $p_x>0$ for energy $E/\mu=20$ and
$\lambda=8$. The cross-section is presented in the
$(\theta_z,{\rm I}_z)$-plane, where $\sqrt{\lambda}\omega_0{\rm I}_z=
p_z^2/2m+(m\lambda\omega_0^2/2)z^2$,
$\arctan\theta_z=m\sqrt{\lambda}\omega_0 z/p_z$. The action variable
${\rm I}_z$ is given in units of $2\mu/\omega_0$. Frequency ratios
$\Omega_x:\Omega_z$ of resonances are given on the right hand margin together
with the resonance actions determined from perturbation theory. For
resonances existing in doublets related by symmetries only one member
of the doublet is shown for clarity. The 5:14 resonance could not be detected
numerically.

\noindent
{\bf Fig.~2.} Poincar\'e surface of sections as in fig.1 for $E/\mu=100$ for
the dynamics with the condensate in Thomas-Fermi approximation (upper part)
and for the condensate with boundary layer of thickness $l_H/\rho_{TF}=0.1$,
continuous up to and including the third derivative (lower part)

\noindent
{\bf Fig.~3.} Poincar\'e surface of sections as in figs.1,2 but plotted in the
$(z,p_z)$-plane. $z,p_z$ are plotted in units $\sqrt{2\mu/m\omega_0^2},
\sqrt{2m\mu}$. The value of $E/\mu$ is 100 (upper left), 20 (upper right),
10 (lower left), and 2 (lower right).

\noindent
{\bf Fig.~4.} Poincar\'e surface of sections as in fig.3,
but for finite angular momentum $L_z = \mu/\omega_0$.
Again the values of $E/\mu$ are 100 (upper left), 20 (upper right),
10 (lower left), and 2 (lower right).

 
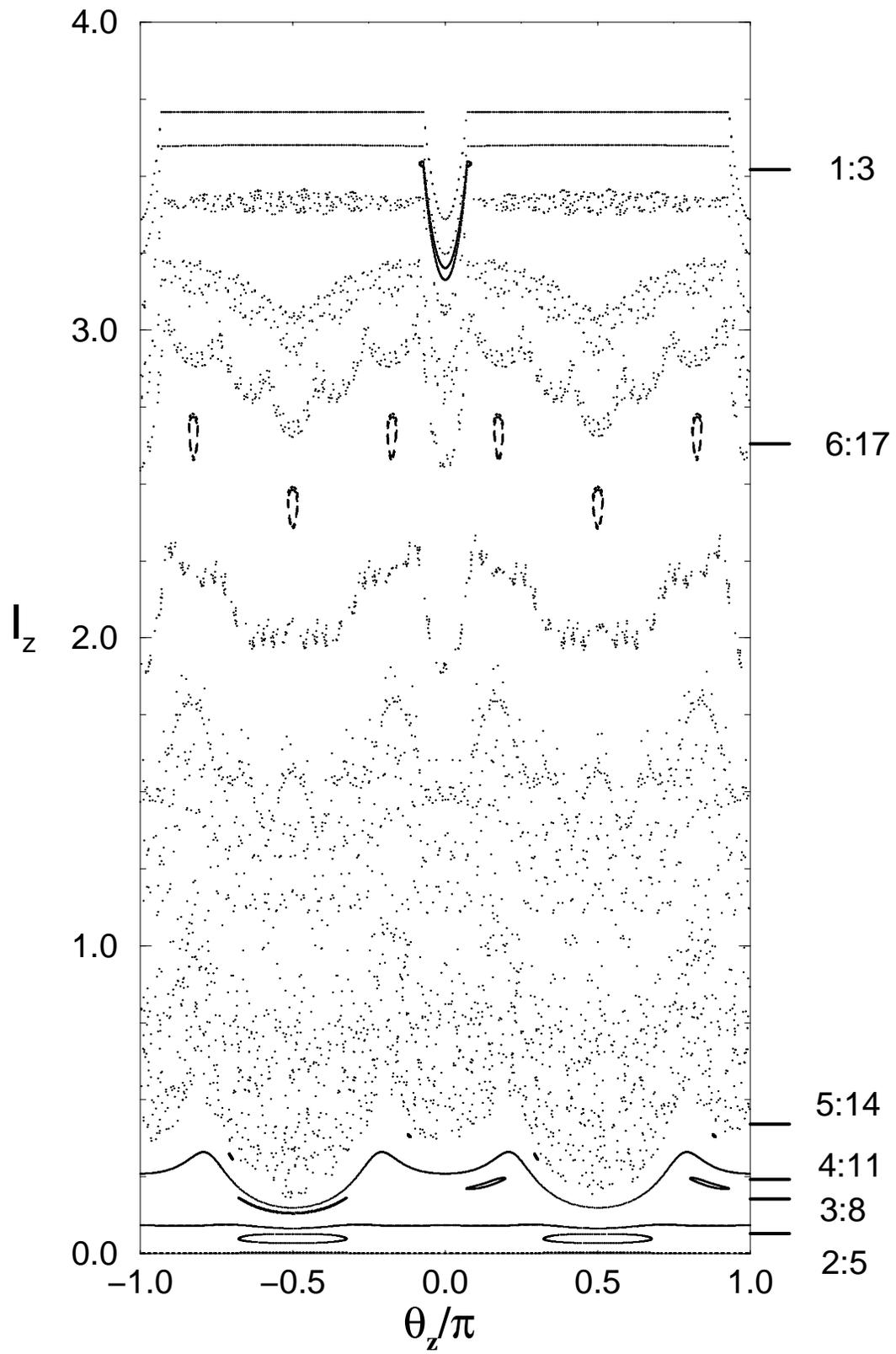
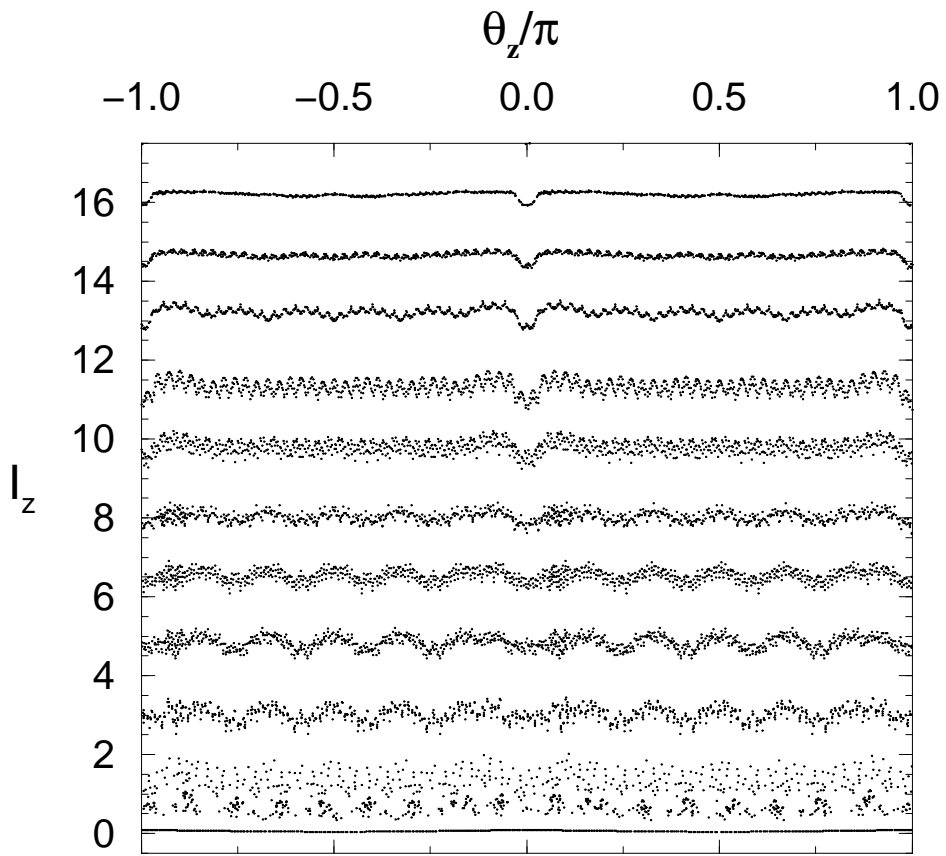
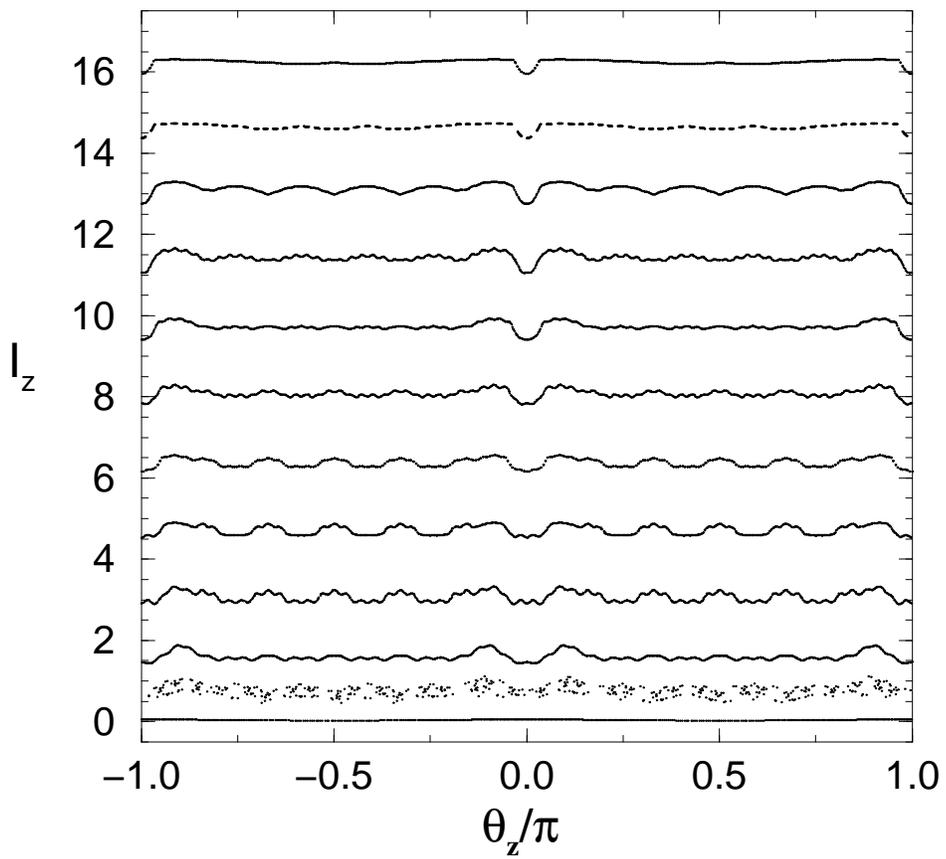
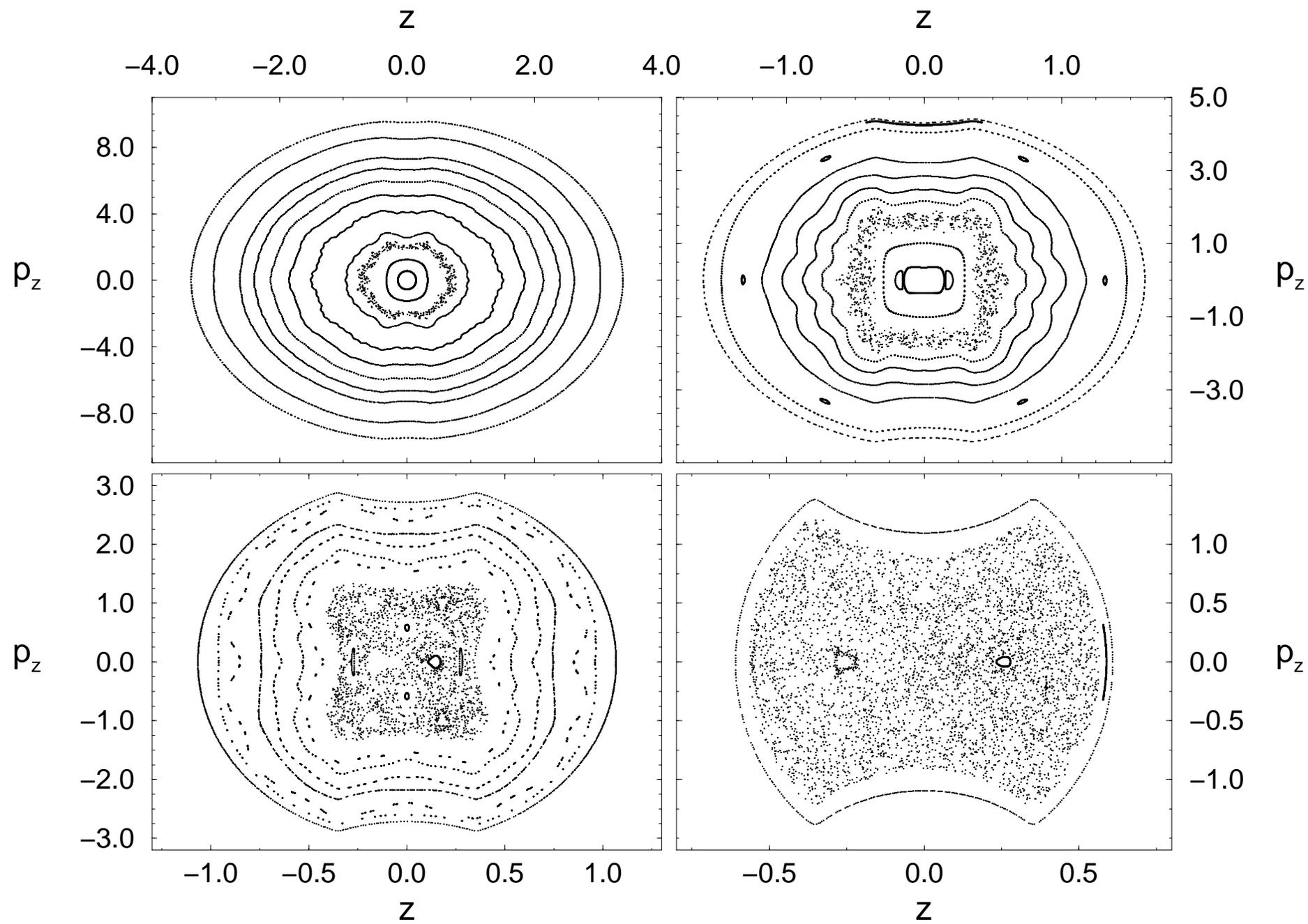
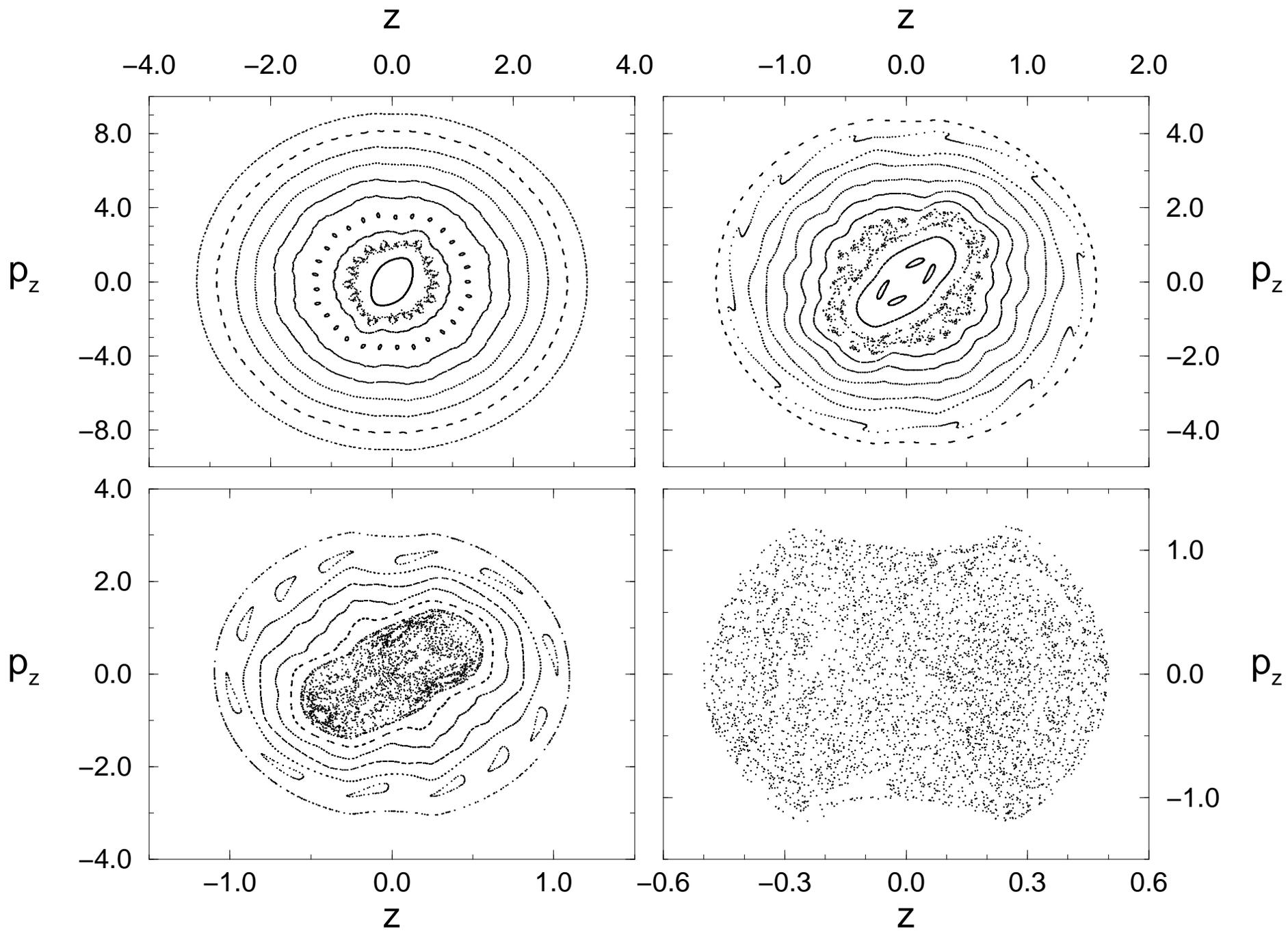
\end{document}